\documentclass{article}
\usepackage{amssymb,amsmath,amsfonts,amsthm,epsfig,subfig,caption}
\def\pmb#1{\setbox0=\hbox{$#1$}%
  \kern-.025em\copy0\kern-\wd0
  \kern.05em\copy0\kern-\wd0
  \kern-.025em\raise.0433em\box0}
\def\pmbs#1{\setbox0=\hbox{$\scriptstyle #1$}%
  \kern-.0175em\copy0\kern-\wd0
  \kern.035em\copy0\kern-\wd0
  \kern-.0175em\raise.0303em\box0}
\def\be{\begin{equation}}
\def\ee{\end{equation}}
\def\bea{\begin{eqnarray}}
\def\eea{\end{eqnarray}}
\def\lb{\label}

\def\r{\ref}

\def\Sigp{\Sigma_{+}}
\def\Sigm{\Sigma_{-}}
\def\Sigc{\Sigma_{\times}}

\def\Nm{N_{-}}
\def\Nc{N_{\times}}

\def\Om{\Omega}

\def\ptl{\partial}

\def\hsp5{\hspace{5mm}}
\newcommand{\sfrac}[2]{{\textstyle{#1\over#2}}}
\def\case#1/#2{\textstyle\frac{#1}{#2}}

\begin{document}

\begin{center}
{\Large\bf Periodic boundary conditions and $G_{2}$ cosmology}
\vspace{.3in} \\ 
{\bf A A Coley}, 
\\Department of Mathematics \& Statistics, Dalhousie University,\\
Halifax, Nova Scotia, Canada B3H 3J5
\\Email: acoley@dal.ca
\vspace{.1in}
\\ {\bf W C Lim},
\\Department of Mathematics, University of Waikato, Private Bag 3105, Hamilton 3240, New Zealand
\\Email: wclim@waikato.ac.nz
\vspace{.1in}
\vspace{0.2in}

\end{center}

\begin{abstract}
In the standard  concordance cosmology the spatial curvature is assumed to be constant and zero (or at least very small). In particular, in numerical computations of the structure of the universe using N-body simulations, exact periodic
boundary conditions are assumed which constrains the spatial curvature. In order to confirm this qualitatively, we numerically evolve a 
special class of spatially inhomogeneous $G_2$ models with both periodic initial data and non periodic initial data using zooming techniques.
We consequently demonstrate that in these models periodic initial conditions do indeed suppress the growth of the  spatial curvature as the models evolve away from their initial isotropic and spatially homogeneous state, thereby verifying that the spatial curvature is necessarily very small in standard cosmology.

\end{abstract}

\section{Introduction}

Cosmology is concerned with the large scale behaviour of the Universe within General Relativity (GR) \cite{ColeyEllis}.
The {\em{cosmological principle}} then implies that on the
largest scales the Universe can be accurately modeled by an exact solution to Einstein’s field equations (EFE) which
is both spatially homogeneous and isotropic leading to a (background) Friedmann-Lemaıtre-Robertson-Walker
(FLRW) model (with constant spatial curvature) with the cosmological constant, $\Lambda$, representing dark energy 
(or the so-called $\Lambda$CDM standard concordance cosmology, where CDM denotes cold dark matter). Inflation in the early universe  is often regarded as a part of the standard model, in which 
the background spatial curvature, characterized by the normalized curvature parameter, is predicted to be negligible \cite{Martin}. Regardless of whether inflation is included
as part of the standard model or not, spatial curvature is implicitly assumed to be zero.

An important goal in cosmology is to understand the origin of the structure of the universe. Assuming that cosmic structure grew out of small initial fluctuations, we can investigate
their evolution on sufficiently large scales using linear perturbation theory (LPT), where the spatially inhomogeneous perturbations {\em{live}} on the flat FLRW background spacetime. Inflation then provides
a causal mechanism for generating primordial cosmological perturbations via 
quantum fluctuations in the inflaton field, which subsequently act as seeds for the anisotropies in the cosmic microwave
background (CMB) \cite{PlanckA6} and give rise to the large scale structure of our universe. At late times and scales much smaller than the Hubble scale fluctuations of the cosmic density are not necessarily small, and LPT is not sufficient
and clustering needs to be treated non-linearly. Usually this is studied with (non-relativistic)
N-body simulations. Recently  non-linear perturbations have been investigated at second-order
and non perturbative relativistic effects have been studied computationally
\cite{ColeyEllis}.

Although standard cosmology has been very successful in describing current observations, there are a number of possible
anomalies \cite{Buchert}, including the tension between the recent determination of the local value of the
Hubble constant based on direct measurements of supernovae \cite{Riess} and the value derived from the most recent CMB data \cite{PlanckA6}, as well as various self-inconsistencies \cite{Valentino}.
In addition, the values of cosmological parameters inferred from the CMB perhaps suggest
a number of directional dependences (e.g., possible greater than anticipated bulk flows on large scales and {\em{dark flow}} \cite{Kashlinsky}).

In principle, since the Universe is not isotropic or spatially homogeneous on small local scales, the
effective smoothed out macroscopic gravitational FE on large scales should be obtained by averaging the EFE of GR.
The averaging of the EFE
for local inhomogeneities can lead to important  backreaction effects on the average evolution
of the Universe \cite{Bolejko}, and can significantly affect precision cosmology at the level of 1 \% \cite{ColeyEllis}.

\subsection{Spatial curvature}

As noted above, in standard cosmology the spatial curvature is assumed to be constant and zero (or at least perturbatively small). {\footnote{By {\it{global}} constant spatial curvature we essentially mean the value of $k$ in the background FLRW spacetime (and hence the initial conditions). By the {\it{local}} value of the spatial curvature we primarily mean the local value obtained later in the paper by the numerical integration across the region of simulation for different specific values of $x_{zoom}$. In the later numerical simulations we also discuss an averaged spatial curvature at a given time.}}
However, there is no completely independent constraint that gaurentees a
value for the magnitude of the effective normalized spatial curvature $\Omega_k$ of less than about 1 \%.
Moreover, any possible such small non-zero measurement of $\Omega_k$ can perhaps be taken to
indicate that the assumptions in
the standard model are not valid. We also note that spatial curvature is,
in general, evolving in relativistic cosmological models \cite{Bolejko}.

In order to develop models to be utilized for cosmological predictions and comparison with observational data, it is necessary to make assumptions.
However, it is crucial to check how the assumptions
affect the results. More importantly,
we can only validate the consistency of assumptions and we cannot rule out alternative explanations. The assumption of a FLRW
background on cosmological scales presents a number of problems \cite{ColeyEllis}. For example, the assumption
that there exists a 1+3 spacetime split and a global time and a background inertial (Gaussian normal) coordinate
system  over a complete
Hubble scale {\em{background}} patch in the standard model leads to the constraint that the spatial
curvature (and the vorticity) must be sufficiently small. In addition, any assumptions of exact periodic
boundary conditions (PBC) (appropriate on scales comparable to the homogeneity scale) necessarily imply that
the global spatial curvature is exactly zero \cite{ColeyEllis} (since in the actual standard model the Universe is taken to
be simply connected), and thus any suitable approximation will
amount to $\Omega_k$ being less than the perturbation (e.g., LPT) scale. We shall return to this point below.

There are also assumptions related to the weak field approach, the actual applicability of perturbation theory and
Gaussian initial conditions, that imply spatial curvature is neglected. It is also asserted that cosmological
backreaction can be neglected, but in LPT the initial fluctuations are usually taken to be Gaussian, which implies that at
the linear level all averages are zero by construction. In addition, we should be aware that any intuition based
on Newtonian theory may be misplaced; e.g., the average spatial curvature of voids and clustered matter
(with negative and positive curvature, respectively) is not necessarily zero within GR.
Thus, within standard cosmology the spatial curvature is assumed to be zero, or at least very small and at
most first order in terms of the perturbation approximation, in order for any subsequent analysis to be
valid. Any prediction larger than this indicates an inconsistency in the approach. In this sense the standard model
cannot be used to predict a small spatial curvature.

It has recently been re-emphasized that, in an era of precision cosmology,  any physical approximations (including, for example, the use of LPT), in
standard cosmological models should be analysed to check if they are
sufficiently accurate for comparison with observations used to draw  conclusions about the actual properties of our Universe \cite{Giblin}.

\subsection{Periodic boundary conditions} 

In the general inhomogeneous cosmological case, the formulation of the evolution eqns. consists of utilizing global Cartesian coordinates and  metric variables, where initial conditions on the metric are specified. 
In the actual N-body simulations
a cell of length L (a fraction of the Hubble scale) is taken and the system is integrated where the PBC at L is applied
at all times (and is used at every step to compute spatial derivatives near the edges of the cell). Hence spatial curvature is held zero on the boundary, and in this sense the growth of spatial curvature is suppressed, relatively speaking.
It is of importance to verify numerically that initial PBC constrain the spatial curvature to be zero at the boundary for all time, which is maintained at zero even below numerical errors that begin to grow. 
 
In addition,  if in N-body simulations the cell is restricted to a finite size L, then periodicity implies that longer wavelength inhomogeneities than L are effectively neglected. 
This will have an effect on the numerical simulations,
which is a general concern, especially when the large-scale cutoff is indeed below the Hubble scale as will be necessary if small ($\sim Mpc$) scales are to be resolved. (There will be a cutoff in practice in the power spectrum of the inhomogeneities immediately from the initial conditions.)

Finally, it is not absolutely clear whether periodic initial conditions guarantee periodic evolution for all times. There is no
general mathematical theorem on the preservation of discrete symmetries (that periodicity implies; however, see \cite{Friedrich}).
But note that  if it were true that periodic initial conditions are propagated, then it follows that the spatial curvature is indeed constrained at L for all times.

In the few current relativistic cosmological simulations, the spatial curvature can have large local fluctuations but remains small when averaged over large scales. It becomes extremely small,   $\Omega_R  \sim 10^{-8}$,  when averaged over the whole periodic box of the simulation \cite{MPL19}.
Work is in progress for a more in-depth investigation of average spatial curvature from numerical relativity, along with the constraints due to the PBC, in a more frame-independent formalism  \cite{MacphersonMourier}.
However, the preliminary evidence is that these
numerical solutions have an exceptionally and unnaturally small spatial curvature in a neighbourhood of the periodic  boundary, which affects the average spatial curvature in the whole cell, and can perhaps be ascribed to the use of PBC.

In different formulations of the EFE  (e.g., the $G_2$ case below), it is difficult to
compare with the standard numerical results of the full 3D code and the usual N-body simulations. Indeed, 
since the EFE are hyperbolic, and depending on the particular formulation of
the evolution equations, once the initial conditions are fully specified it is 
not clear whether the application of spatial PBC are even required (beyond the specification of the initial data that is).

\section{Spatially inhomogeneous  models}

The FLRW cosmology (or Friedmann–Lemaıtre (FL) model as we will refer to as here when regarding the solution as a dynamical  equilibrium state) is an expanding exactly spatially
homogeneous and spatially isotropic universe, which is believed to describe the
Universe on a sufficiently large spatial scale, at least since the time of last scattering. However, the observable part of the Universe is not exactly spatially homogeneous and isotropic on
any spatial scale. From a practical point of view, we are interested in 
studying models that are {\em{close to FL}}
in some appropriate dynamical sense, perhaps 
using deviations from an FL model by applying LPT. But it is not known how reliable the linear theory is and, moreover,
in using it we are a priori excluding the possibility of uncovering important non-linear effects.

Therefore, it
is of interest to consider (more) general spatially inhomogeneous  models (than FL)
in order to investigate the constraints that observations
impose \cite{WE}. Indeed, it could be argued that
we should not study exotic ideas until conventional GR (with inhomogeneities) has been fully explored. In the first instance, it is perhaps advisable to investigate
a special inhomogeneous model where some analytical and qualitative insights are possible.
In this paper we wish to
{\em{study the effect of spatial curvature and PBC
numerically in the special case of $G_2$ models.}}

\subsection{$G_2$ models}

The simplest spatially inhomogeneous cosmological models have two commuting Killing vector fields (i.e., models admitting a 2-parameter Abelian isometry group acting transitively on spacelike 2-surfaces), which consequently have a single
degree of spatial inhomogeneity. This class of models, which are referred to briefly as
$G_2$ cosmologies, are governed by the EFE which constitute partial differential equations (PDE)
in two independent variables  \cite{ElstUggla}. A lot of the research to date has focused on the so-called Gowdy vacuum spacetimes
\cite{Gowdy}, because such models are much more tractable than non-vacuum ones and the existence of compact space sections facilitates numerical simulations since the problem of boundary conditions
at spatial infinity is avoided. Recently, a plane-parallel model (or wall universe) \cite{grasso},  in which light rays pass through a series of plane-symmetric
perturbations around a FLRW background, was used to study the back-reaction from the small-scale inhomogeneities 
\cite{Villa, Adame}.

Belinskii, Khalatnikov and Lifshitz (BKL) \cite{BLK}  have conjectured that within GR
the approach to the generic spacelike  singularity to the past is vacuum dominated (assuming $p < \rho$),
local, and oscillatory (labeled ‘Mixmaster’). Localilty implies that the contribution of terms in the
evolution eqns. with spatial derivatives can be neglected. Numerical studies of the asymptotics
of the simplest inhomogeneous vacuum Gowdy models demonstrate that on
approach to the singularity, spiky structures form \cite{Berger}. These spikes become  narrower as the
singularity is approached. Studies of $G_2$ and more general models have produced numerical evidence that
the BKL conjecture generally holds except possibly at isolated points where spiky structures form \cite{Andersson,Lim}.

We are particularly interested in late time evolution here. We shall concentrate on $G_2$ cosmologies with a perfect
fluid matter content, and especially dust \cite{thesis}. At any instant of time, the state of a $G_2$ cosmology is described by a finite-dimensional dynamical
state vector of functions of the spatial coordinate. That is, the dynamical state space of $G_2$
cosmologies is a function space and is thus infinite-dimensional. The evolution of a $G_2$ cosmology
is consequently described by an orbit in this infinite-dimensional dynamical state space. In a $G_2$ cosmology,  at each point
there exists a preferred timelike 2-space that is orthogonal to the $G_2$–orbits. Therefore, there is an
infinite family of geometrically preferred timelike congruences and gauge freedom in the choice of the
orthonormal frame (and associated with
the choice of the local coordinates).
Following \cite{orthonormal,ElstUggla}, we shall utilize the orthonormal frame formalism adapted to the $G_2$ orbits.
Such a dynamical formulation has proved effective in studying $G_2$ cosmologies. Here we define scale-invariant dependent variables by normalisation with the area expansion rate of
the $G_2$–orbits, in order to obtain the evolution eqns. as a system of PDE in first-order symmetric
hyperbolic (FOSH) format. This FOSH format also provides a natural framework for numerical studies.


\subsection{$G_2$ evolution equations in timelike area gauge}


The geometry of the general $G_2$ class of spacetimes is given in \cite{Lim}, where all of the metric functions depend only on the time
coordinate $t$ and the spatial coordinate $x$. 
We will display the $G_2$ evolution system in terms of the timelike area gauge in the Gowdy subcase $\Sigma_2 = 0$ (with unit lapse, thereby
defining $t$ time). 
We utilize $\beta$-normalized variables \cite{ElstUggla,thesis} in the orthonormal frame formalism \cite{orthonormal}, where the $t$
and $x$ subscripts denote partial differentiation. 
We also assume dust ($\gamma = 1$) and a single non-zero tilt component, $v$.  We are particularly interested in future dynamics close to 'FL'. We also assume that the cosmological constant is zero, so that we have an
unconstrained evolution system in FOSH format \cite{ElstUggla}:

\bea
\lb{tlareae11dot}
\ptl_{t}E_{1}{}^{1}
& = & (q+3\Sigp)\,E_{1}{}^{1} \\
\lb{tlareasigmdot}
\ptl_{t}\Sigm + E_{1}{}^{1}\,\ptl_{x}\Nc
& = & (q+3\Sigp-2)\,\Sigm + 2\sqrt{3}\,\Sigc^{2}
- 2\sqrt{3}\,\Nm^{2} \\
\lb{tlareancdot}
\ptl_{t}\Nc + E_{1}{}^{1}\,\ptl_{x}\Sigm
& = & (q+3\Sigp)\,\Nc \\
\lb{tlareasigcdot}
\ptl_{t}\Sigc - E_{1}{}^{1}\,\ptl_{x}\Nm
& = & (q+3\Sigp-2-2\sqrt{3}\Sigm)\,\Sigc - 2\sqrt{3}\,\Nc\,\Nm \\
\lb{tlareanmdot}
\ptl_{t}\Nm - E_{1}{}^{1}\,\ptl_{x}\Sigc
& = & (q+3\Sigp+2\sqrt{3}\Sigm)\,\Nm + 2\sqrt{3}\,\Sigc\,\Nc
\eea
\be
\lb{tlareaomdot}
(\ptl_{t}
+ v\,E_{1}{}^{1}\,\ptl_{x})\,\Om
+ \Om E_{1}{}^{1}\,\ptl_{x}v
 =  2\,\Om \ [(q+1)
- \sfrac{1}{2}\,(1-3\Sigp)\,(1+v^{2}) - 1] 
\ee
\be
\lb{tlareavdot}
(\ptl_{t} + v\,E_{1}{}^{1}\,\ptl_{x})\,v =  (1-v^{2})\,
[\ 3(\Nc \Sigm - \Nm \Sigc) + \sfrac{3}{2}\Om v - (1-3\Sigp)\,v]
\ee
where
\be
\lb{ca:q3sigp}
(q+3\Sigp) \equiv  2 - \sfrac{3}{2}\,(1-v^{2})\,\Om \ 
\ee
and
\be
\lb{ca:sigpt}
\Sigp  \equiv  \sfrac{1}{2}\,(1-\Sigm^{2}-\Nc^{2}-\Sigc^{2}
-\Nm^{2}-\Om). \\
\ee

\noindent
Note that in the present case we have from Eqs. (\r{ca:q3sigp}) and
(\r{ca:sigpt}) that 
\be
	q = \frac12 + \frac32( \Sigm^2 + \Nc^2 + \Sigc^2 + \Nm^2) + \frac32v^2\Omega \geq \frac12,
\ee
which, together with
\be
	\partial_t \beta = -(q+1) \beta,
\ee
guarantees that  $\beta$ is single signed and is monotone (for small $\epsilon$ -- see below).
The (negative) spatial curvature ($-\Omega_k$) is defined via
\be
\lb{ca:om}
\Omega_k \equiv \Nc^{2} + \Nm^{2}.
\ee
The relation of the variables above with metric components is given by (see (6.39) and (6.40) of \cite{thesis}, or Appendix A.3 of \cite{ElstUggla}):
\be
	ds^2 = \beta^{-2} [ -dt^2 + (E_1{}^1)^{-2} dx^2 ] + e^{2t} [ e^{P(t,x)} (dy + Q(t,x) dz)^2 + e^{-P(t,x)} dz^2 ]
\ee
\bea
	( \Sigm ,\ \Nc ) & = & \frac{1}{2\sqrt{3}} ( \partial_t,\ - E_1{}^1 \partial_x ) P
\\
	( \Sigc ,\ \Nm ) & = & \frac{1}{2\sqrt{3}} e^P ( \partial_t, E_1{}^1 \partial_x ) Q.
\eea

\subsubsection{Initial data}

Since we have an unconstrained FOSH system no constraints need be satisfied and we freely
specify  $\{E_{1}{}^{1},~\Sigm,~\Nc,~\Sigc,~\Nm,~\Om, v \}$ at $t=0$ on $\{-L \leq x \leq L \}$.

\noindent We shall consider initial data close to a (flat) FL model of the form (for small $\epsilon \sim 10^{-4}$):

$\{E_{1}{}^{1} = 1 +  \epsilon^2 {\tilde E}$ \},

$\{\Sigm = \epsilon{\tilde \Sigma}_-, \Nc = \epsilon{\tilde N}_\times, \Sigc = \epsilon{\tilde \Sigma}_\times,   \Nm =  \epsilon{\tilde N}_-  \} $,

$\{ \Om = 1 - \epsilon^2{\tilde \Om}, v = \epsilon^2 {\tilde v} \}$.

\noindent Note that it follows that

$\{q = \frac{1}{2} + \epsilon^2{\tilde q}, \Sigp = \epsilon^2 {\tilde \Sigma}_+, \Omega_k =   \epsilon^2 {\tilde \Omega}_k \}$.

The initial data is given by (for example) $\epsilon{\tilde \Sigma}_-$ at $t = 0$ (etc.). Since shell crossings can occur in dust models,
we either cease integration once any singular behaviour develops or we can include a small fiducial
pressure in the models.

\subsubsection{Linear regime}

For small $\epsilon$, we integrate eqn. (1) to obtain:

\be
\lb{ca:Eoneone}
{\tilde E} = e^{\frac{t}{2}} +  o(\epsilon^2),
\ee
where we have normalized ${\tilde E}$ to be unity at $t = 0$.
In the linear regime we have that $e^{\frac{t}{2}}$ is small (which for $\epsilon \sim 10^{-4}$
corresponds to $t < 10$). Eqns. (\ref{tlareaomdot},\ref{tlareavdot}) also
constitute o($\epsilon^2$) corrections to zeroth order evolution equations. In the linear regime,
$\Omega_k \sim o(\epsilon^2)$.

\subsubsection{Shear and curvature}
Eqns.  (\ref{tlareasigmdot}) -- (\ref{tlareanmdot}) represent the first order evolution eqns. for the normalized shear and curvature variables
with second order corrections. We immediately see the growing modes for the normalized curvature
variables $\sim e^{\frac{t}{2}}$
and the decaying modes for the normalized shear variables $\sim e^{-\frac{3t}{2}}$,
 corresponding to the
familiar eigenvalues $\{\frac{1}{2}, -\frac{3}{2} \}$ for the flat FL ”saddle point” solution [so that the growth of the shear
is suppressed relative to that of the spatial curvature in the initial linear regime].

Solving eqns. (\ref{tlareasigmdot}) - (\ref{tlareanmdot}) we obtain:

\bea
\lb{tlareasigmdot2}
{\tilde \Sigma}_- & = &  e^{-\frac{3t}{2}}[\sigma_-   + \epsilon{\bar \Sigma}_-],\\
\lb{tlareancdot2}
{\tilde N}_\times & = & e^{\frac{t}{2}}[\nu_\times + \epsilon{\bar N}_\times],\\
\lb{tlareasigcdot2}
{\tilde \Sigma}_\times &= &   e^{-\frac{3t}{2}}[\sigma_\times +  \epsilon{\bar \Sigma}_\times], \\
\lb{tlareanmdot2}
{\tilde N}_- & = & e^{\frac{t}{2}}[\nu_- + \epsilon{\bar N}_-],
\eea
where $\{\sigma_-, \sigma_\times, \nu_-, \nu_\times \}$ are slowly varying (and, for example, 
$\ptl_{x}\Sigm \sim (\epsilon e^{\frac{t}{2}}) {\bar \Sigma}_-{}^{\prime}$,
$\ptl_{x}\Sigc \sim (\epsilon e^{\frac{t}{2}}) {\bar \Sigma}_\times{}^{\prime}$,
where a prime denotes $\ptl_{x}$). In the regime in which the shear is sub-dominant, these quantities are approximately
constant.

\subsubsection{Spatial curvature}
From eqns.  (\ref{tlareancdot}) and 
(\ref{tlareanmdot}), to $o(\epsilon)$ eqn. (\ref{ca:om}) becomes:
\be
\lb{ca:om25}
\ptl_{t}\Omega_k = \Omega_k,
\ee
so that $\Omega_k$ remains small (and of second order initially). More precisely, to $o(\epsilon^4)$:
\be
\lb{ca:om27}
\ptl_{t}\Omega_k = (\epsilon e^{\frac{t}{2}})^2[(\nu_-^{2} + \nu_\times ^{2}) + \epsilon
(\nu_\times {\bar{N}_\times} + \nu_- {\bar{N}_-}) +4 \sqrt{3}(\epsilon e^{\frac{t}{2}})e^{-2t}
(\sigma_- \nu_- + \sigma_\times \nu_{\times}) +  (\epsilon e^{\frac{t}{2}})e^{-2t}
({\nu_-}{ \bar \Sigma}_\times{}^{\prime} + \nu_\times{\bar \Sigma}_-{}^{\prime}).
\ee

Clearly to second order we duplicate eqn. (\ref{ca:om25}); the leading order correction to this eqn. comes from the
term $\epsilon (\nu_- {\bar{N}_\times} + \nu_\times {\bar{N}_-})$, which corresponds to eqn. (\ref{ca:om27}) to next order. The remaining terms are of order
$o(\epsilon e^{-2t})$ and $o(\epsilon^{2})$. For early times $\ptl_{t}\Omega_k  > 0$, and so the magnitude of the spatial curvature grows. The
sign of the next order terms is not necessarily positive.

\subsubsection{Boundary conditions}

We can consider PBC at $X = \pm L$.
Any periodic initial data for any variable $X$ on $[-L, L]$, for $L$ sufficiently large, can be Fourier decomposed. Note
that at $t = 0$ the average values are $\langle X \rangle = c_0$, which is usually taken to be zero, and $\langle X^{\prime} \rangle = 0$. We are also interested in non periodic initial data,
since in the formulation of $G_2$ models here PBC are not necessary

\subsubsection{Numerical methods: Zooming}

One numerical method for resolving small scale structure is adaptive mesh refinement (AMR). However,
if we know beforehand the location of the structure, we need not use AMR.
Because we wish to numerically study  non-PBC, we shall consequently use zooming techniques, 
in which we use a
coordinate system adapted to the structure that is under investigation. In particular, we can choose a coordinate system that shrinks exponentially with time, in which the
new coordinates zoom in on the worldline, and where the constant rate of focus $A$ is controlled  
($A = 1$ is the natural choice, which ensures a good excision boundary throughout the simulation) \cite{Lim}. 
The numerical grid ends at a fixed coordinate value $X = L$. Ordinarily, that would call for a
boundary condition at $L$, but we can use the method of excision, which can be applied to any hyperbolic
equations where the outer boundary is chosen so that all causal influences are outgoing. In that case the equations of motion are simply
implemented at the outer boundary; no boundary condition is needed (or even allowed).

\section{Results}

\begin{figure}[t]
	\centering
	\includegraphics[width=0.9\textwidth]{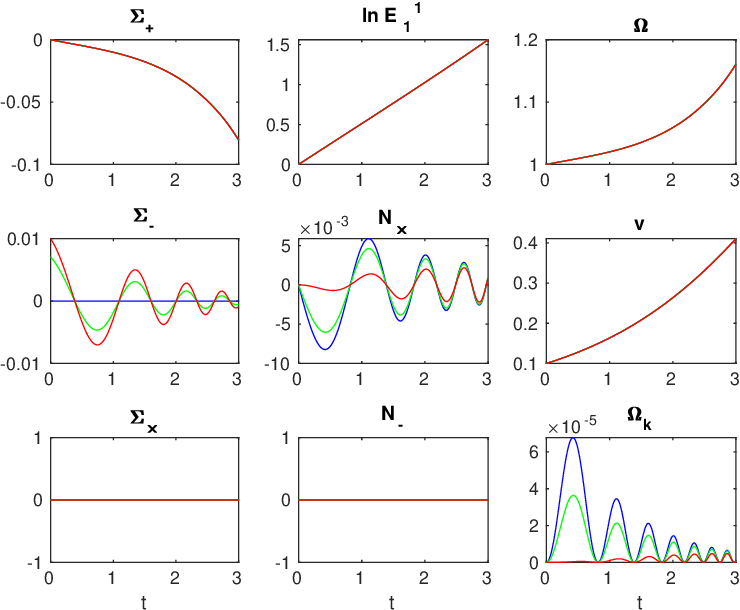}
	\caption{In Figure~\ref{fig:1a}--\ref{fig:1c} we consider three numerical runs for the initial conditions 
(a) $\Sigma_- = 0.01 \sin \pi x$ 
(b) $\Sigma_- = 0.01 \sin \pi x + 0.01$ 
(c) $\Sigma_- = 0.01 \sin \pi x + 0.01x$ 
(each zooming in at $x_\text{zoom} = 0, 0.25, 0.5$, plotted in blue, green, red, respectively) in the diagonal $G_2$ models. 
The domain is $x - x_\text{zoom} \in [-10, 10]$, with $t$ running from $t = 0$ to $t = 3$.
The initial conditions can be interpreted as a perturbed open FL model (small $\Sigma_-$) that is also close to a flat FL model (small $v$). 
For all initial conditions $v$ grows from 0.1 to 0.4, taking the solution further away from the flat FL model. 
The growth of the spatial curvature is displayed in the final frame in each figure for each initial condition.}	
	\label{fig:1a}
\end{figure}

\begin{figure}[t]
        \centering
        \includegraphics[width=0.9\textwidth]{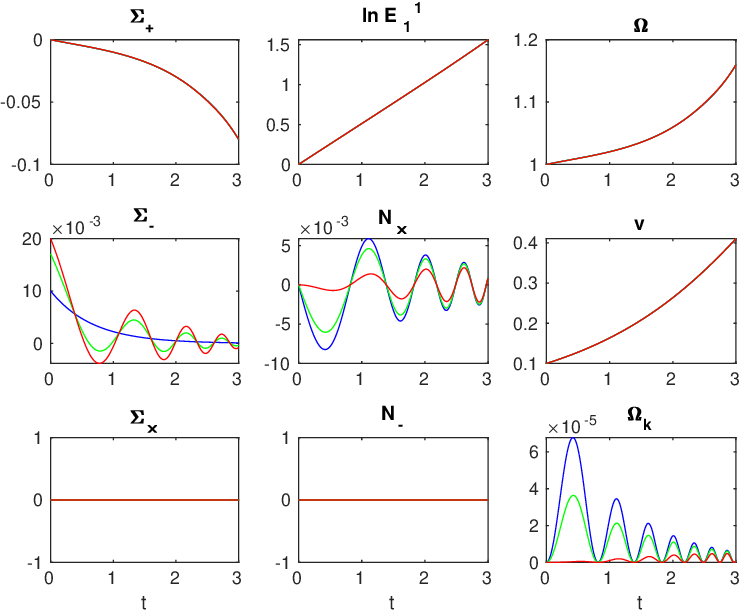}
	\caption{(See Figure~\ref{fig:1a} caption.)}
	\label{fig:1b}
\end{figure}

\begin{figure}[t]
	\centering
        \includegraphics[width=0.9\textwidth]{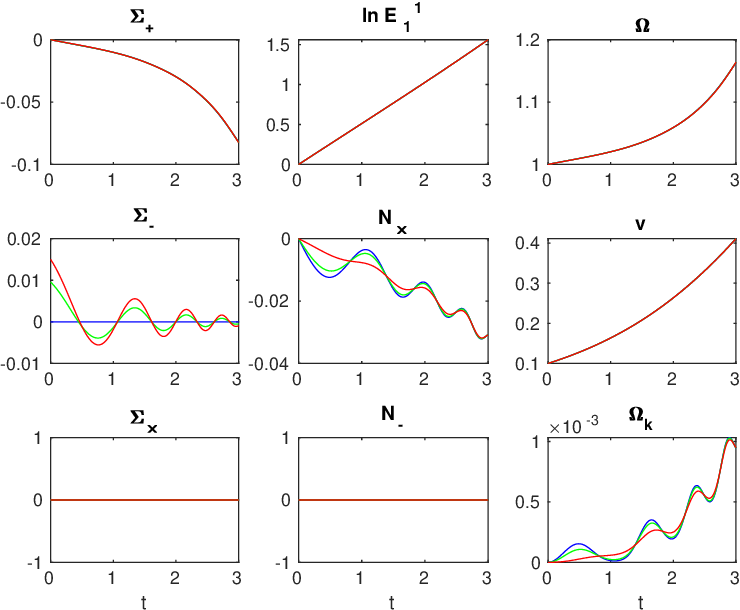}
	\caption{(See Figure~\ref{fig:1a} caption.)}
	\label{fig:1c}
\end{figure}

In Figures~\ref{fig:1a}--\ref{fig:1c} we present three numerical simulations
for the $G_2$ dust evolution system in terms of the timelike area gauge
and using 
$\beta$-normalized variables with the initial conditions
$\{ E_{1}{}^{1} = 1, \Sigm = {\tilde{\Sigma}_-}, \Nc = 0, \Sigc = 0, \Nm =  0,
\Om = 1 -  {\tilde{\Sigma}_-}^2, v = 0.1 \}$ (which actually correspond to {\em{diagonal}} $G_2$ models, with $\Sigma_\times = N_- = 0$ at all times) and with
(a) ${\tilde{\Sigma}_-} = 0.01 \sin \pi x$ 
(b) ${\tilde{\Sigma}_-} = 0.01 \sin \pi x + 0.01$ 
(c)  ${\tilde{\Sigma}_-}= 0.01 \sin \pi x + 0.01x$ 
(each zooming in at $x_\text{zoom} = 0, 0.25, 0.5$, plotted in blue, green, red, respectively). The 
domain is $x - x_\text{zoom}$ in $[-10, 10]$ with zooming, with $t$ running from $t = 0$ to $t = 3$. 
The initial conditions can be
interpreted as a perturbed open FL model (small $\Sigma_-$) that is also close to a flat FL model (small $v$).  The numerical results 
demonstrate that for all  initial conditions $ v$
grows from 0.1 to 0.4, taking the solution further (drifting) away from flat FL
(regarded as a {\em{saddle point}} in the dynamical state space), regardless of the
periodicity or aperiodicity  of the initial conditions.

From the final frame in the Figs. it is clearly demonstrated that $\Omega_k$ is suppressed in Fig. 1(a)
with periodic initial conditions, and the growth of $\Omega_k$ is evident when 
periodicity is broken in Fig. 1(c), which
can be attributed to the non-periodic term $0.01x$ in the initial conditions for $\Sigma_-$. Note that this non-periodic term has a  very large wavelength.  However, the simulations Fig. 1(b) show that the presence of a constant term 0.01 in $\Sigma_-$ is found to have a  negligible effect on the spatial curvature  $N_\times$ and $\Omega_k$, since the large background homogeneous term in the shear in this model (i.e., $\Sigma_- = 0.01 \sin \pi x + 0.01$, which is still periodic) does not pass into $N_\times$. Additional simulations with 
various initial conditions were not found to change the qualitative behaviour of these
results.

\section{Discussion}

We have shown that in this particular $G_2$ model periodic initial conditions do indeed suppress the  spatial curvature, and hence we have verified that their use implies a small spatial curvature a priori. More comprehensive numerical analysis is possible but, as we noted above, this may not directly improve the qualitative conclusions regarding more general 3D simulations.

Indeed, in the $G_2$ FOSH formulation of the evolution eqns., the variables are not metric variables and consequently the initial conditions are not specified on the metric functions. Hence the meaning of periodic initial data has a different meaning to that in the usual standard sense.  Indeed, in this formulation once the initial conditions are fully specified the application of spatial boundary conditions are not even required. It is difficult to compare directly with the results obtained in the usual full 3D numerical
simulations. Hence it is important to repeat this type of analysis in general cosmological simulations \cite{MacphersonMourier}.

If the periodic initial conditions are maintained, the only arbitrarily large wavelength modes allowed are the background homogeneous terms, and they are only allowed in the shear components (which are time derivatives of the metric) but not in the spatial curvature components (which are spatial derivatives of the metric).  
That is, the spatially homogeneous growing mode is turned off, and long-wavelength modes which behave like Bianchi dynamics, leaving only short-wavelength modes which grow slower (if they grow at all). 
Periodicity generally disallows arbitrarily large wavelengths in the spatial curvature.

In the analysis there is a potential gauge issue which concerns the choice of the timelike area gauge (which affects the normalization of the independent variables such as  $\Omega_k$). (The usual gauge issues when using LPT are not relevent here in the numerical simulations). The emphasis in this short paper is to qualitatively illustrate the relative growth of $\Omega_k$ in the evolutions without  PBC compared to those with PBC. Other numerical investigations were undertaken, but none challenge this conclusion and none, in our opinion, really led to any further insights. So although the gauge affects the actual quantitative results, it does not affect the qualitative results (at least according to our computations). In particular, the behaviour of the growth of $\Omega_k$ (in the figures) is typical as found across the region of simulation, and in other timelike area gauge numerical simulations, indicating that the relative average spatial curvature is increasing in the models without PBC. In addition, comparison of $\Omega$ and $\Omega_k$ in the figures (where the influence of the normalization factor is then minimised) illustrates that the qualitative growth of $\Omega_k$ in the evolutions without PBC compared to those with PBC does not depend on the choice of the timelike area gauge.

\section*{Data availability statement}

The data that support the findings of this study are openly available at the following URL/DOI:
\begin{verbatim}
www.math.waikato.ac.nz/~wclim/pbcpaper2023_source_files/. 
\end{verbatim}

\section*{Acknowledgement}

AAC would like to acknowledge NSERC for financial assistance. We would like to thank Pierre Mourier
and Hayley Macpherson for helpful comments and private communications.




\begin{thebibliography}{99}


\bibitem{ColeyEllis} A. A. Coley and G. F. R. Ellis,  Class. Quant. Grav. {\bf{37}} 013001 (2020) [arXiv:1909.05346].

\bibitem{Martin} J. Martin [arXiv:1902.05286].



\bibitem{PlanckA6} Planck Collaboration: Planck 2018 results,  Astronomy \& Astrophysics {\bf{641}} A6  \& A7 (2020); Y. Akrami et al. [arxiv:1807.06205 \& 06209 \& 06211]; 



\bibitem{Buchert} T. Buchert et al., Int. J. Mod. Phys. D {\bf{25}} 1630007 (2016).


\bibitem{Riess} A. G. Riess et al., A Comprehensive Measurement of the Local Value of the Hubble Constant with 1 km/s/Mpc
Uncertainty from the Hubble Space Telescope and the SH0ES Team [arXiv:2112.04510]; A. G. Riess et al., Astrophys. J. {\bf{861}} 126 (2018) [arxiv:1804.10655].


\bibitem{Valentino} E. Di Valentino et al.,  Class. Quant. Grav.  {\bf{38}} 153001 (2021); M. Asgari et al.,
Astronomy \& Astrophysics {\bf{645}}  A104 (2021); T. M. Abbott et al., Phys. Rev. D {\bf{105}} 023520 (2022).

\bibitem{Kashlinsky} A. Kashlinsky et al., Astrophysical Journal Letters {\bf{686}} L49 (2008); C. Howlett et al., Monthly Notices of the Royal
Astronomical Society {\bf{515}} 953 (2022).


\bibitem{Bolejko} K. Bolejko, Phys. Rev. D {\bf{97}} 103529 (2018);
D. L. Wiltshire, New J. Phys. {\bf{9}} 377 (2007) \& Class. Quant. Grav. {\bf{28}} 164006 (2011); T. Buchert et al.,
Class. Quant. Grav. {\bf{32}} 215021 (2015); A. Coley, N. Pelavas and R. Zalaletdinov, Phys. Rev Letts. {\bf{95}} 151102 (2005).

\bibitem{Giblin} J. T. Giblin Jr, J. B. Mertens, G. D. Starkman and C. Tian,
Phys. Rev. D {\bf{99}} 023527 (2019) [arXiv:1810.05203 [astro-ph.CO]]; P. K. Aluri et al., Is the observable universe consistent with the cosmological principle? [arXiv:2207.05765].

\bibitem{Friedrich} H. Friedrich and A. Rendall. "The Cauchy problem for the Einstein equations." Einstein’s Field Equations and Their Physical Implications: Selected Essays in Honour of Jürgen Ehlers (Springer, Berlin Heidelberg,  pp 127-223, 2000).




\bibitem{MPL19} H. Macpherson, J. Price and P. Lasky, Phys. Rev. D
{\bf{99}} 3522 (2019).

\bibitem{MacphersonMourier} H. Macpherson and P. Mourier, in preparation. 



\bibitem{WE} J. Wainwright and  G. F. R. Ellis (eds)  Dynamical Systems in Cosmology (Cambridge  Cambridge
University Press, 1997).

\bibitem{ElstUggla}  H. van Elst, C. Uggla, and J. Wainwright, Class. Quant. Grav. {\bf{19}} 51 (2002) [arXiv:gr/qc/0107041].


\bibitem{Gowdy} R. H. Gowdy, Phys. Rev. Lett. {\bf{27}} 826  (1971) \& Ann. Phys. (N.Y.) {\bf{83}} 203 (1971).

\bibitem{grasso} M. Grasso, E. Villa, M. Korzyński and S. Matarrese,
Isolating non-linearities of light propagation in inhomogeneous cosmologies
[arxiv/2105.04552].

\bibitem{Villa} E. Villa, S. Matarrese, and D. Maino, JCAP {\bf{2011}} 024 (2011).

\bibitem{Adame} J. Adamek, E. Di Dio, R. Durrer and M. Kunz, Phys.
Rev. D {\bf{89}} 063543 (2014).



\bibitem{BLK} 
V. A. Belinskiı, I. M. Khalatnikov, and E. M. Lifshitz, 
Adv. Phys., {\bf{12}} 185 (1963) \& {\bf{19}} 525 (1970) \&  {\bf{31}} 639 (1982).




\bibitem{Berger} B. K. Berger and V. Moncrief, Phys. Rev. D {\bf{48}} 4676 (1993);
B. K. Berger and D. Garfinkle, Phys. Rev. D {\bf{57}} 4767 1998);
B. K. Berger, J. Isenberg and M. Weaver, Phys. Rev. D {\bf{64}} 084006 (2001);
D. Garfinkle, Phys. Rev. Lett. {\bf{93}} 161101 (2004).



\bibitem{Andersson}  L. Andersson, H. van Elst, W. C. Lim and C. Uggla, Phys. Rev. Lett. {\bf{94}} 051101 (2005).



\bibitem{Lim}  W. C. Lim, L. Andersson, D. Garfinkle and F. Pretorius,  Phys. Rev. D {\bf{79}} 123526 (2009) [arXiv: 0904.1546].

\bibitem{thesis}  W. C. Lim, Ph.D. thesis, University of Waterloo 2004) [gr-qc/0410126].

\bibitem{orthonormal} H. van Elst  and C. Uggla,  Class. Quant.
Grav. {\bf{14}} 2673  (1997).


\end{thebibliography}
\end{document}